\documentclass{jpsj-suppl}
\usepackage{txfonts} 

\makeatletter
\def\simleq{\mathrel{\mathpalette\gl@align<}}
\def\simgeq{\mathrel{\mathpalette\gl@align>}}
\def\gl@align#1#2{\lower.6ex\vbox{\baselineskip\z@skip\lineskip\z@
     \ialign{$\m@th#1\hfill##\hfil$\crcr#2\crcr\sim\crcr}}}
\makeatother

\newcommand{\bec}[1]{\mbox{\boldmath $#1$}}

\newcommand{\Luscher}{L\"uscher}

\title{Towards Lattice QCD Baryon Forces at the Physical Point: First Results}

\author{%
Takumi~\textsc{Doi},$^1$ 
Sinya~\textsc{Aoki},$^{1,2,3}$
Shinya~\textsc{Gongyo},$^{1,2}$
Tetsuo~\textsc{Hatsuda},$^{1,4}$
Yoichi~\textsc{Ikeda},$^1$
Takashi~\textsc{Inoue},$^{1,5}$
Takumi~\textsc{Iritani},$^{1,6}$
Noriyoshi~\textsc{Ishii},$^{1,7}$
Takaya~\textsc{Miyamoto},$^{1,2}$
Keiko~\textsc{Murano},$^{1,7}$
Hidekatsu~\textsc{Nemura},$^{1,3}$
and
Kenji~\textsc{Sasaki}$^{1,3}$
}

\inst{%
$^1$Theoretical Research Division, Nishina Center, RIKEN, Wako 351-0198, Japan\\
$^2$Yukawa Institute for Theoretical Physics, Kyoto University, Kyoto 606-8502, Japan\\
$^3$Center for Computational Sciences, University of Tsukuba, Ibaraki 305-8571, Japan\\
$^4$Kavli IPMU (WPI), The University of Tokyo, Chiba 277-8583, Japan\\
$^5$Nihon University, College of Bioresource Sciences, Kanagawa 252-0880, Japan\\
$^6$Department of Physics and Astronomy, Stony Brook University, Stony Brook, New York 11794-3800, USA\\
$^7$Research Center for Nuclear Physics (RCNP), Osaka University, Osaka 567-0047, Japan\\
}

\email{doi@ribf.riken.jp}

\recdate{December 13, 2015}

\abst{%
Lattice QCD calculations of baryon forces are performed
for the first time with (almost) physical quark masses.
$N_f = 2+1$ dynamical clover fermion gauge configurations are generated 
at the lattice spacing of $a \simeq 0.085$ fm on a $(96 a)^4 \simeq (8.2 {\rm fm})^4$ lattice
with quark masses corresponding to $(m_\pi, m_K) \simeq (146, 525)$ MeV.
Baryon forces are calculated 
using the time-dependent HAL QCD method.
In this report, we study $\Xi\Xi$ and $NN$ systems
both in $^1S_0$ and $^3S_1$-$^3D_1$ channels,
and the results for the central and tensor forces 
as well as phase shifts in the $\Xi\Xi$ $(^1S_0)$ channel are presented.
}

\kword{nuclear forces, hyperon forces, lattice QCD}

\begin{document}
\maketitle

\section{Introduction}
\label{sec:intro}

The determination of baryon forces 
based on the fundamental theory, Quantum Chromodynamics (QCD),
is one of the most challenging issues in nuclear physics.
There have been first-principles lattice QCD calculations
for baryon forces,
using the \Luscher's finite volume method~\cite{Luscher:1990ux, Yamazaki:2015nka}
or 
the (time-dependent) HAL QCD method~\cite{Ishii:2006ec, Aoki:2012tk, Ishii:hyp2015}.
The latter method is particularly useful since 
one can extract
energy-independent (non-local) 
potentials from 
Nambu-Bethe-Salpeter (NBS) correlators
without the issue associated with the ground state saturation
on a lattice~\cite{Aoki:2012tk}.
(For detailed comparison between the \Luscher's method and the HAL QCD method, 
see Ref.~\cite{Iritani:2015dhu}.)
The actual lattice simulations, however, have been limited to unphysically
heavy quark masses
due to the lack of computational resources.

Under these circumstances, 
we have launched a new project 
which aims at the lattice calculations of baryon forces 
with physically light quark masses on a large lattice volume,
exploiting 
the supercomputers such as Japanese flagship K computer.
The lattice QCD simulations for the baryon forces play a complementary role
to the experiments in the sense that the 
precision of the former becomes better for larger values of strangeness $|S|$,
while the situation is opposite in the latter.
In this paper, we present the latest status report in this (on-going) project~\cite{Doi:lat2015}.
We study $\Xi\Xi$ and $NN$ systems both in $^1S_0$ and $^3S_1$-$^3D_1$ channels,
and obtain the central and tensor forces.
The results for other hyperon forces are given in Refs.~\cite{Ishii:hyp2015, Sasaki:hyp2015}.

\section{Lattice QCD setup}
\label{sec:setup}

$N_f = 2+1$ gauge configurations are generated 
on a $96^4$ lattice
using the clover fermion with stout smearing.
Using K computer, about 2000 trajectories are generated,
and preliminary studies show that 
the lattice volume amounts to $(8.2 {\rm fm})^4$ with the lattice spacing $a \simeq 0.085$ fm,
and ($m_\pi, m_K) \simeq (146, 525)$ MeV~\cite{Ishikawa:2015rho}.
For further details on the gauge configuration generation,
see Ref.~\cite{Ishikawa:2015rho}.

We consider all 52 channels relevant to two-octet baryon forces in parity-even channel.
Corresponding NBS correlators are calculated with the same quark masses used in the configuration generation,
where wall quark source with Coulomb gauge fixing are employed.
The computational cost for NBS correlators is significantly reduced by the unified contraction algorithm~\cite{Doi:2012xd}.
Combined with the efficient solver for quark propagators~\cite{Boku:2012zi},
the performance efficiency for the total measurement computations
achieves $\sim$~17\% on 2048 nodes ($\times$ 8 cores/node) of K computer,
corresponding to $\sim$~45 TFlops sustained.
We pick 1 configuration per each 10 trajectories,
and
the rotation symmetry is used to increase the statistics.
The total statistics used in this report amounts to
203 configurations $\times$ 4 rotations $\times$ 20 wall sources.

Baryon forces are determined from NBS correlators in the time-dependent HAL QCD method
in $^1S_0$ and $^3S_1$-$^3D_1$ channels.
It is guaranteed that obtained potentials are faithful to the phase shifts
by construction~\cite{Aoki:2012tk}.
NBS correlators are evaluated at several $t$ (Euclidean temporal distance) and 
we examine the trade-off between systematic and statistical errors:
the results from larger $t$ suffer from smaller systematics due to inelastic states on a lattice, 
while statistical fluctuations become larger.
We perform the velocity expansion~\cite{Aoki:2012tk} in terms of 
the non-locality of potentials,
and obtain the leading order potentials, i.e., central and tensor forces.
In this preliminary analysis shown below, 
the term which corresponds to the relativistic effects is omitted for simplicity~\cite{Doi:lat2015}.

\begin{figure}[t]
\begin{minipage}{0.48\textwidth}
\begin{center}
%
\includegraphics[angle=0,width=0.85\textwidth]{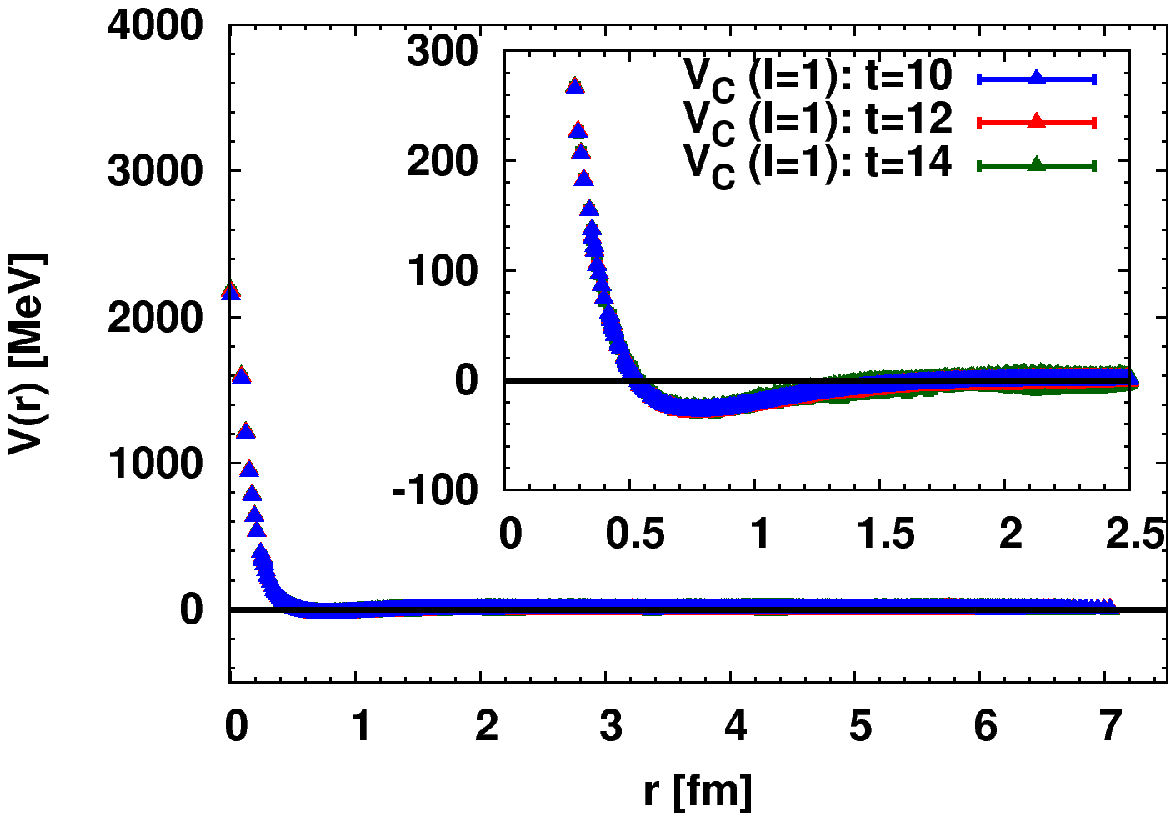}
%
\caption{
\label{fig:pot:XiXi:1S0:cen}
$\Xi\Xi$ central force $V_C(r)$ in $^1S_0$ $(I=1)$ channel
obtained at $t = 10, 12, 14$.
}
\end{center}
\end{minipage}
\hfill
\begin{minipage}{0.48\textwidth}
\begin{center}
%
\includegraphics[angle=0,width=0.85\textwidth]{figs/XiXi/phase.spline.t_011-015.eps.hyp15_proc}
%
\caption{
\label{fig:phase:XiXi:1S0:cen}
$\Xi\Xi$ scattering phase shifts in $^1S_0$ $(I=1)$ channel
obtained at $t = 10, 12, 14$.
}
\end{center}
\end{minipage}
\vspace*{-3mm}
\end{figure}

\section{Results}
\label{sec:results}
\vspace*{-2mm}

\begin{figure}[t]
\begin{minipage}{0.48\textwidth}
\begin{center}
%
\includegraphics[angle=0,width=0.85\textwidth]{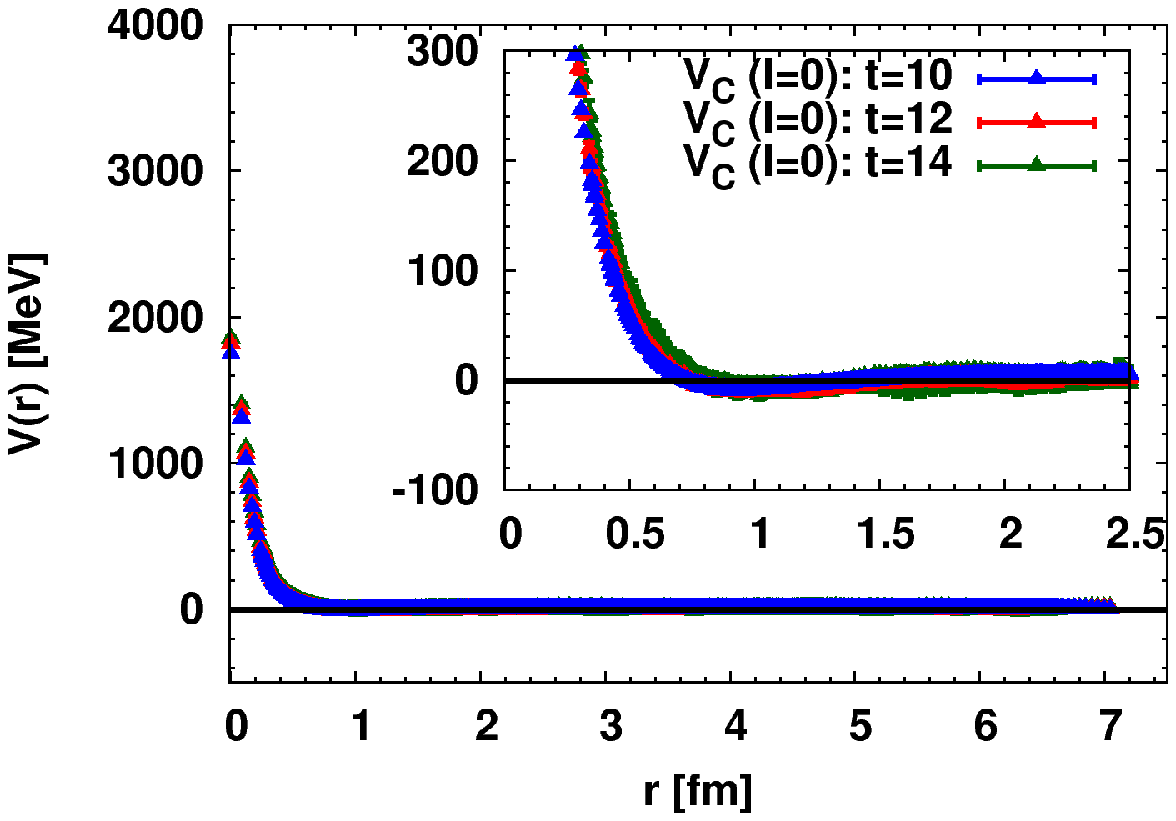}
%
\caption{
\label{fig:pot:XiXi:3S1:cen}
$\Xi\Xi$ central force $V_C(r)$ in $^3S_1$-$^3D_1$ $(I=0)$ channel
obtained at $t = 10, 12, 14$.
}
\end{center}
\end{minipage}
\hfill
\begin{minipage}{0.48\textwidth}
\begin{center}
%
\includegraphics[angle=0,width=0.85\textwidth]{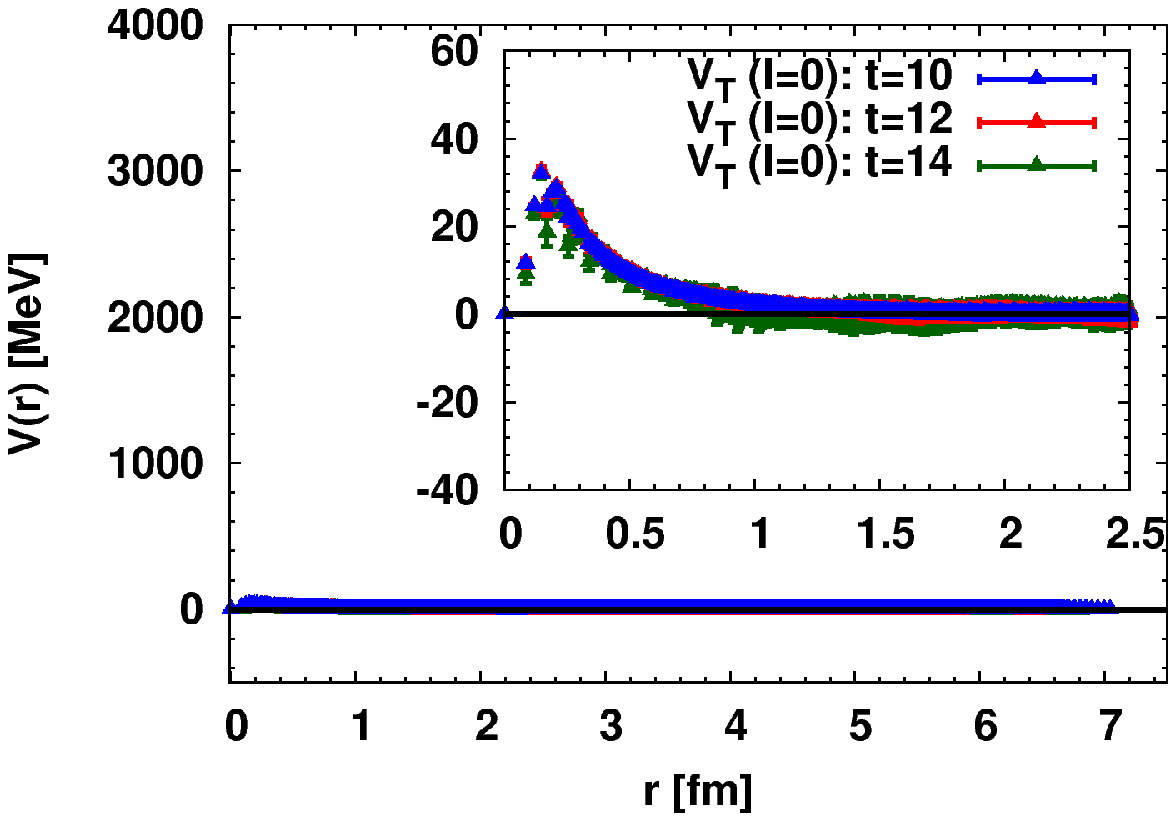}
%
\caption{
\label{fig:pot:XiXi:3S1:ten}
$\Xi\Xi$ tensor force $V_T(r)$ in $^3S_1$-$^3D_1$ $(I=0)$ channel
obtained at $t = 10, 12, 14$.
}
\end{center}
\end{minipage}
\end{figure}

\subsection{$\Xi\Xi$ systems}
\label{subsec:XiXi}

We first consider the $\Xi\Xi$ system in $^1S_0$ (iso-triplet) channel.
This channel belongs to the 27-plet 
in flavor SU(3) classification
as does the $NN$ system in $^1S_0$ channel.
Therefore, 
the $\Xi\Xi (^1S_0)$ interaction serves as a good ``doorway'' to probe 
the $NN (^1S_0)$ interaction,
where the signal in the former is much cleaner than the latter on a lattice.
In addition,
since the strong attraction in the $NN (^1S_0)$ channel makes a ``dineutron'' nearly bound,
it has been attracting interest whether 
the 27-plet interaction with the SU(3) breaking effects
forms a bound $\Xi\Xi (^1S_0)$ state or not~\cite{Haidenbauer:2014rna, Rijken:hyp2015}.

In Fig.~\ref{fig:pot:XiXi:1S0:cen},
we show the lattice QCD results
for the central force 
in the $\Xi\Xi (^1S_0)$ channel.
We observe a clear signal of 
the mid- and long-range attraction as well as the repulsive core at short-range.
Within statistical fluctuations,
the results are found to be consistent with each other in the range $t = 10-14$,
which suggests that the contaminations from inelastic excited states are suppressed
and higher-order terms in the velocity expansion are small.

We fit the potential with a spline curve and 
calculate the phase shifts by solving the Schr\"odinger equation in the infinite volume.
Shown in Fig.~\ref{fig:phase:XiXi:1S0:cen}
are the obtained phase shifts in terms of the center-of-mass energy.
The results indicate that the interaction is strongly attractive at low energies 
while it is not sufficient to form a bound $\Xi\Xi (^1S_0)$ state.
Further studies with larger statistics are currently underway.
It is also desirable to examine this observation by, e.g.,
heavy-ion collision experiments.

We then consider the $\Xi\Xi$ system in $^3S_1$-$^3D_1$ (iso-singlet) channel.
This channel belongs to the 10-plet in flavor SU(3),
a unique representation with hyperon degrees of freedom.
By solving the coupled channel Schr\"odinger equation with 
NBS correlators, we determine the potentials. 
In Figs.~\ref{fig:pot:XiXi:3S1:cen} and ~\ref{fig:pot:XiXi:3S1:ten},
we show the central and tensor forces, respectively,
where the tensor operator is defined as 
$\hat{S}_{12}(\bec{r}) = 3(\bec{\sigma_1}\cdot\bec{r})(\bec{\sigma_2}\cdot\bec{r})/r^2 - (\bec{\sigma_1}\cdot\bec{\sigma_2})$.
For the central force, 
we observe the strong repulsive core,
which is in accordance with the quark Pauli blocking effect~\cite{Aoki:2012tk, Oka:1986fr}.
There also exists a weak attraction at mid-range but 
larger statistics are necessary for the conclusive argument.
We observe that the $\Xi\Xi$ tensor force (Fig.~\ref{fig:pot:XiXi:3S1:ten})
has opposite sign and is much weaker than 
the $NN$ tensor forces (Fig.~\ref{fig:pot:NN:3S1:ten}).
This could be understood by the phenomenological 
one-boson exchange potentials ($\pi$ and $\eta$)
with flavor SU(3) meson-baryon couplings.
We also note that both of central and tensor forces in $^3S_1$-$^3D_1$ channel
are found to be somewhat more sensitive to the change of $t$ compared to those in $^1S_0$ channel.
Studies with larger $t$ and larger statistics are in progress.

\subsection{$NN$ systems}
\label{subsec:NN}

\begin{figure}[t]
\begin{minipage}{0.48\textwidth}
\begin{center}
%
\includegraphics[angle=0,width=0.85\textwidth]{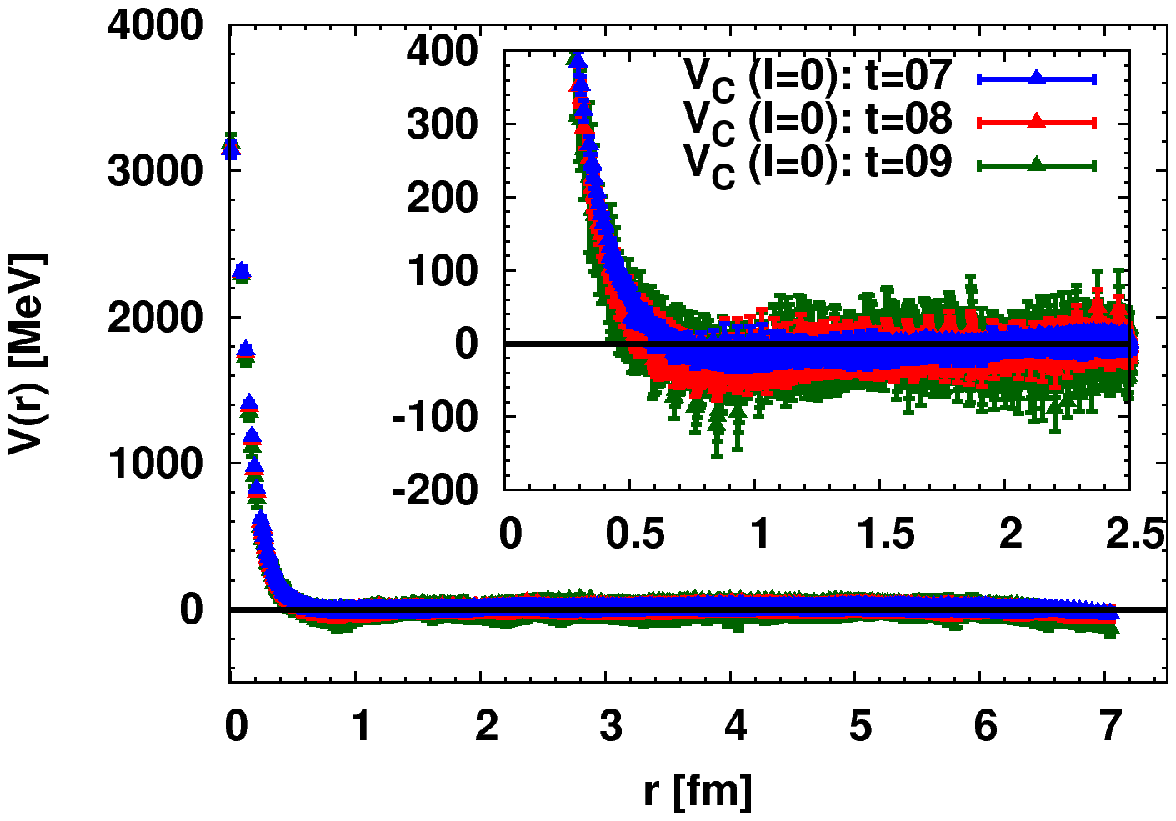}
%
\caption{
\label{fig:pot:NN:3S1:cen}
$NN$ central force $V_C(r)$ in $^3S_1$-$^3D_1$ $(I=0)$ channel
obtained at $t = 7, 8, 9$.
}
\end{center}
\end{minipage}
\hfill
\begin{minipage}{0.48\textwidth}
\begin{center}
%
\includegraphics[angle=0,width=0.85\textwidth]{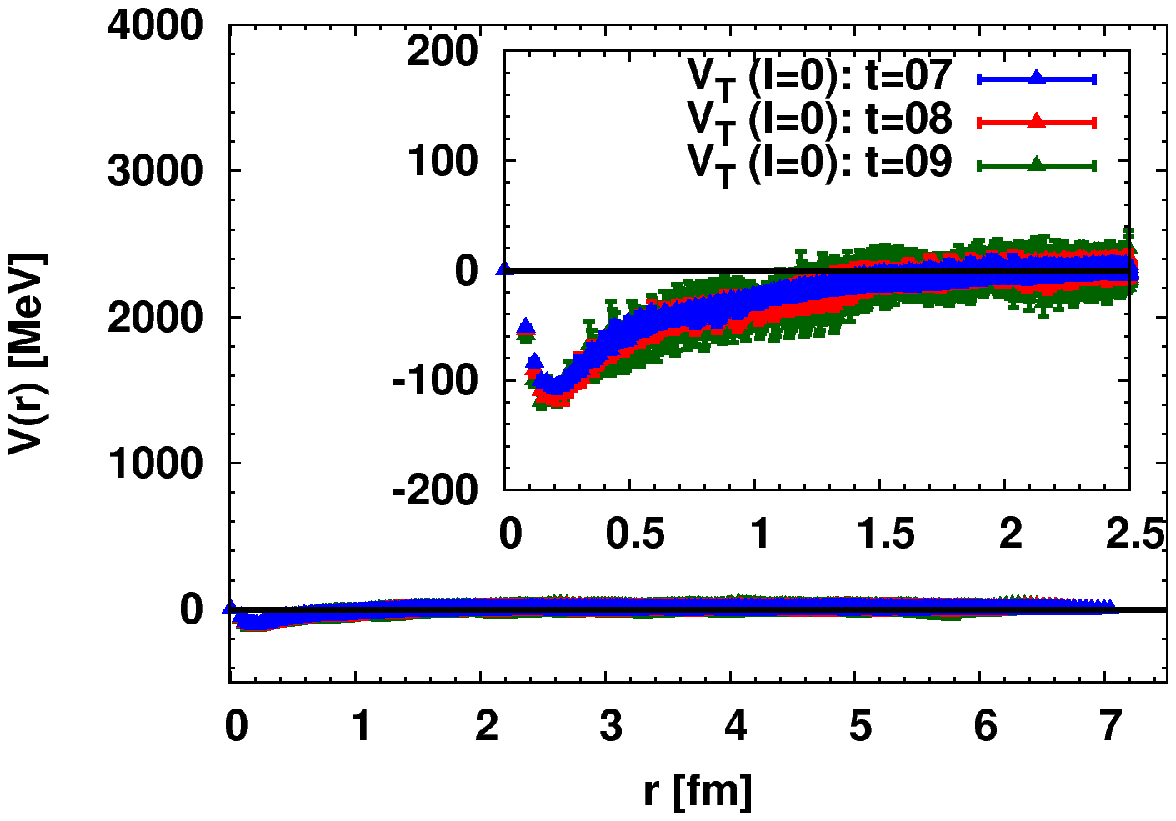}
%
\caption{
\label{fig:pot:NN:3S1:ten}
$NN$ tensor force $V_T(r)$ in $^3S_1$-$^3D_1$ $(I=0)$ channel
obtained at $t = 7, 8, 9$.
}
\end{center}
\end{minipage}
\vspace*{-5mm}
\end{figure}

Let us first consider
$NN$ in $^3S_1$-$^3D_1$ (iso-singlet) channel.
This channel belongs to the 10$^*$-plet in flavor SU(3).
Shown in Figs.~\ref{fig:pot:NN:3S1:cen} and \ref{fig:pot:NN:3S1:ten}
are the central and tensor forces, respectively.
The results suffer from larger statistical errors than those of $\Xi\Xi$,
even though we take smaller $t$ than the case of $\Xi\Xi$.
However, in the central force, the repulsive core at short-range is obtained 
and it is also encouraging that mid- and long-range attraction tends to appear as 
we take larger $t$.
In addition,
it is 
remarkable that the strong tensor force with the long-range tail is clearly visible,
qualitatively in accordance with phenomenological potentials and/or the 
structure of one-pion exchange potential (OPEP).
Compared to the lattice tensor forces obtained with heavier quark masses,
the range of interaction is found to be longer.
Since it is the tensor force which plays the most crucial role in 
the binding of deuteron,
this result is very intriguing.

We also study the central force in $^1S_0$ channel.
Qualitatively similar results as those of the central force in $^3S_1$-$^3D_1$ channel are obtained:
Although the results suffer from large statistical fluctuations,
we observe a clear signal of the repulsive core at short-range,
as well as the tendency that mid- and long-range attraction tends to appear as we take larger $t$~\cite{Doi:lat2015}.

To obtain more quantitative results for nuclear forces, 
it is desirable to take larger $t$ by increasing the statistics, which is currently underway.

\section{Summary}
\label{sec:summary}
\vspace*{-2mm}

We have presented the first lattice QCD calculations
of baryon forces 
which employ almost physical quark masses.
$N_f = 2+1$ dynamical clover fermion gauge configurations have been generated 
at the lattice spacing of $a \simeq 0.085$ fm on a $(96 a)^4 \simeq (8.2 {\rm fm})^4$ lattice,
where $(m_\pi, m_K) \simeq (146, 525)$ MeV.
Baryon forces have been calculated
using the time-dependent HAL QCD method.

In this report, 
we have shown the preliminary results for $\Xi\Xi$ and $NN$ systems.
In the $\Xi\Xi (^1S_0)$ central force, 
we have observed a strong attraction,
although it is not strong enough to form a bound state.
In the $\Xi\Xi$ ($^3S_1$-$^3D_1$) channel, 
we have observed the strong repulsive core in the central force.
Also a small $\Xi\Xi$ tensor force with opposite sign from the $NN$ tensor force has been found.
Nuclear forces have been studied as well in $^1S_0$ and $^3S_1$-$^3D_1$ channels.
In particular, we have observed a clear signal of the tensor force.
Investigations with larger statistics are under progress.

\vspace*{-1mm}
\section*{Acknowledgments}
\vspace*{-2mm}

The lattice QCD calculations have been performed on the K computer at RIKEN, AICS
(Nos. hp120281, hp130023, hp140209, hp150223),
HOKUSAI FX100 computer at RIKEN, Wako (No. G15023)
and HA-PACS at University of Tsukuba (Nos. 14a-20, 15a-30).
We thank ILDG/JLDG~\cite{conf:ildg/jldg}
which serves as an essential infrastructure in this study.
This work is supported in part by 
MEXT Grant-in-Aid for Scientific Research (15K17667, 25287046, 26400281),
and SPIRE (Strategic Program for Innovative REsearch) Field 5 project.
We thank all collaborators in this project.

\end{document}